\documentclass[preprint,12pt]{elsarticle}

\usepackage{amssymb}

\usepackage{amsmath}
\usepackage{graphicx}
\usepackage{booktabs}
\usepackage{latexsym} 
\usepackage{array}    
\usepackage{multirow}
\usepackage{subcaption}
\usepackage{algorithm}
\usepackage[noend]{algpseudocode}
\usepackage{threeparttable}
\usepackage{paralist}   
\usepackage{xspace}
\usepackage{color}
\usepackage{adjustbox}
\usepackage{balance}
\usepackage{wasysym}
\usepackage{rotating}   
\usepackage{listings}
\usepackage{tabu}
\usepackage{pifont}
\usepackage{framed}
\usepackage[shortlabels]{enumitem}

\usepackage{url}            
\usepackage{hyperref}

\usepackage{microtype}

\usepackage{xpatch}
\makeatletter
\chardef\TPT@@@asteriskcatcode=\catcode`*
\catcode`*=11
\xpatchcmd{\threeparttable}
  {\TPT@hookin{tabular}}
  {\TPT@hookin{tabular}\TPT@hookin{tabu}}
  {}{}
\catcode`*=\TPT@@@asteriskcatcode
\makeatother

\newcommand{\red}[1]{\textcolor[rgb]{1.00,0.00,0.00}{#1}}

\newcommand{\green}[1]{\textcolor[rgb]{0.00,0.60,0.00}{#1}}

\newcommand{\cha}{\red{\ding{55}}\xspace}
\newcommand{\gou}{\green{\ding{52}}\xspace}

\definecolor{wheat1}{rgb}{1.000000,0.905882,0.729412}
\definecolor{LightGray}{rgb}{0.827451,0.827451,0.827451}

\newcolumntype{a}{>{\columncolor{wheat1}}l}

\definecolor{mygreen}{rgb}{0,0.6,0}
\definecolor{mygray}{rgb}{0.5,0.5,0.5}
\definecolor{mymauve}{rgb}{0.58,0,0.82}
\definecolor{darkblue}{rgb}{0.0,0.0,0.6}
\definecolor{maroon}{RGB}{102, 0, 0}
\definecolor{Maroon}{cmyk}{0,0.87,0.68,0.32}
\definecolor{darkred}{RGB}{139, 0, 0}
\definecolor{forestgreen}{RGB}{34, 139, 34}

\lstdefinelanguage{XML}
{
  basicstyle=\ttfamily\small,   
  morestring=[b]",
  moredelim=[s][\color{darkblue}]{<}{\ },
  moredelim=[s][\color{darkblue}]{</}{>},
  moredelim=[l][\color{darkblue}]{/>},
  moredelim=[l][\color{darkblue}]{>},
  morecomment=[s]{<?}{?>},
  morecomment=[s]{<!--}{-->},
  stringstyle=\color{darkred},
  identifierstyle=\color{mymauve}
}

\lstdefinestyle{customJava}{
  breaklines=true,
  keepspaces=true,
  frame=single,
  language=Java,
  showstringspaces=false,
  basicstyle=\footnotesize\ttfamily,
  keywordstyle=\color{blue},
  otherkeywords={+, getIntent},
  numbers=left,
  numbersep=5pt,
  numberstyle=\scriptsize\color{black},
  rulecolor=\color{black},
  stepnumber=1,
  tabsize=2,
  commentstyle=\itshape\color{green!40!black},
  stringstyle=\color{orange},
  emph=[1]  
  {
        do,
        try,
        new,
        catch,
        while,
        SecProvider,
        SecReceiver,
        SecService,
        SecActivity,
        SecSink,
  },
  emphstyle=[1]{\color{darkred}},
  emph=[2]  
  {
        @Override,
  },
  emphstyle=[2]{\color{purple!40!black}},
  belowskip=-1em, 
}

\newif\ifANNOYMIZE
\ANNOYMIZEtrue

\newif\ifACM
\ACMtrue  

\ifACM
\fi

\ifACM
\newcommand{\myfig}{Figure\xspace}
\else
\newcommand{\myfig}{Fig.\xspace}
\fi

\ifACM
\newcommand{\mysec}{\S}
\else
\newcommand{\mysec}{Section\xspace}
\fi

\newcommand{\name}{AGChain\xspace} 
\newsavebox{\bigimage} 

\journal{Journal of Systems Architecture}

\begin{document}

\begin{frontmatter}



\title{\name: A Blockchain-based Gateway for Trustworthy App Delegation from Mobile App Markets}


\author[1]{Mengjie Chen\fnref{label1}}
\fntext[label1]{The work by Mengjie Chen was done while at the Chinese University of Hong Kong.}

\author[2]{Xiao Yi\fnref{label2}}
\fntext[label2]{Mengjie Chen and Xiao Yi are the co-first authors.}

\author[2]{Daoyuan Wu\corref{cor1}}
\cortext[cor1]{Daoyuan Wu is the corresponding author. Contact email: \texttt{dywu@ie.cuhk.edu.hk}.}

\author[3]{Jianliang Xu}
\author[4]{Yingjiu Li}
\author[5]{Debin Gao}

\affiliation[1]{organization={Mask Network},
            city={Shanghai},
            country={China}}
\affiliation[2]{organization={The Chinese University of Hong Kong},
            city={Hong Kong SAR},
            country={China}}

\affiliation[3]{organization={Hong Kong Baptist University},
            city={Hong Kong SAR},
            country={China}}

\affiliation[4]{organization={University of Oregon},
            city={Eugene},
            state={Oregon},
            country={USA}}

\affiliation[5]{organization={Singapore Management University},
            city={Singapore},
            country={Singapore}}

\begin{abstract}


The popularity of smartphones has led to the growth of mobile app markets, creating a need for enhanced transparency, global access, and secure downloading.
This paper introduces \name, a blockchain-based gateway that enables trustworthy app delegation within existing markets.
\name ensures that markets can continue providing services while users benefit from permanent, distributed, and secure app delegation.
During its development, we address two key challenges: significantly reducing smart contract gas costs and enabling fully distributed IPFS-based file storage.
Additionally, we tackle three system issues related to security and sustainability.
    We have implemented a prototype of \name on Ethereum and Polygon blockchains, achieving effective security and decentralization with a minimal gas cost of around 0.002 USD per app upload (no cost for app download).
The system also exhibits reasonable performance with an average overhead of 12\%.

\end{abstract}



\begin{keyword}

Blockchain \sep Smart contract \sep Ethereum \sep IPFS \sep App security



\end{keyword}

\end{frontmatter}


\section{Introduction}\label{sec:intro}

The increasing popularity of smartphones worldwide is driven by the wide range of feature-rich mobile apps available in various app markets.
In addition to official marketplaces like Google Play and the Apple Store, third-party app markets such as Amazon AppStore and Baidu Market have emerged as significant supplements to official app markets.
These third-party markets offer a greater variety of app options for Android users and have gained popularity, particularly in China due to restrictions imposed by the Great Firewall.
However, users' growing concerns regarding security and privacy have created a demand for permanent, distributed, and secure app access in both official and third-party app markets (more details are available in \S\ref{sec:motivate}):

%
%
\begin{description}
  \item[D1:] \textit{Permanent or more transparent app access.}
Firstly, there is a need for enhanced transparency and permanence in app access.
Currently, app markets often lack access to older versions of apps, and developers have the freedom to delist apps at their discretion.
For example, our analysis of 8,359 popular apps downloaded in November 2018 and 2019~\cite{DSDK21} demonstrates that 13.7\% (1,146) of these apps were no longer available after just one year.
This emphasizes the demand for permanent or more transparent app access.
Presently, users must resort to ad-hoc backup solutions~\cite{apprestore, appback} to address this issue.

  \item[D2:] \textit{Global and distributed access.}
Secondly, there is a growing demand for global and distributed access to mobile apps.
Currently, many apps on Google Play and the Apple Store are limited to certain countries, hindering their availability worldwide.
Moreover, Google Play itself is subject to censorship in certain regions, resulting in restricted app access for users in those affected areas.

  \item[D3:] \textit{Secure access.}
Thirdly, our measurement shows that a considerable number of third-party Android app markets lack secure app downloading mechanisms.
For instance, half of the top 14 Chinese app markets, including popular platforms like Baidu and 360, download apps through insecure HTTP connections.
Furthermore, even in markets where secure downloading is available, they generally lack app repackage checking~\cite{DroidMOSS12, Repack19, WardenAttack21} found in official markets. This poses a significant risk to users' apps, as their integrity and security could be compromised.
\end{description}

While individual measures such as VPN usage (for N2) and HTTPS implementation (for N3) may\footnote{Considering that the majority of general users do not have access to paid VPN services and the security status of Chinese app markets is unlikely to change quickly, relying on these technical means may not be as effective as initially assumed.} address some of the demands, they do not adequately resolve the fundamental limitation in terms of scientific research: \textit{the lack of trustworthy app access}.
Developing a decentralized app market from scratch, such as SkyDroid~\cite{SkyDroid}), as similar to existing blockchain-based solutions~\cite{IoTDataBlockchain17, PubChain19, DClaims19, BBox20, DBLP:journals/jpdc/HeFPCTCPNX22}, may appear to be a logical approach to tackle these challenges.
However, this approach introduces a new predicament, as existing markets already contain a substantial number of apps.
Without a sufficient inventory of apps, a decentralized market would be ineffective for end users.

In this paper, we present a novel architecture that integrates the strengths of traditional IT infrastructure with decentralized blockchain technology.
Our proposed solution is \name (\underline{A}pp \underline{G}ateway \underline{Chain}), a blockchain-based gateway that connects end users with existing app markets.
\name provides users with a secure, permanent, and distributed method of app delegation, ensuring trustworthy app downloads.
With \name, users can opt for indirect app downloads, mitigating concerns related to direct downloads from existing markets.
While existing markets continue to provide services, with the exception of delisted apps\footnote{In addition to the apps available on existing markets, AGChain allows users to upload custom apps through a GitHub URL, such as \url{https://github.com/agchain/agchain/blob/main/Test/A.apk}. This feature is specifically designed to accommodate apps that may require payment or could potentially be delisted in the future.}, users can delegate app downloads by inputting the market or custom app URL into \name.
\name then securely retrieves the app from its original market, uploads the raw app file to decentralized storage, and stores the app file index and metadata on the blockchain for future direct downloads through \name.
Once delegated, the app can be permanently and distributedly downloaded from \name by the user and other users.

We utilize smart contracts~\cite{buterin2014next} for implementing our logic on the blockchain and leverage IPFS (Interplanetary File System)~\cite{IPFS14} for decentralized storage.
However, rather than employing them in a conventional manner like other blockchain- and IPFS-based systems~\cite{IoTDataBlockchain17, PubChain19, DClaims19, BBox20}, we address and overcome two previously neglected challenges:
\begin{itemize}
\item 
Firstly, we significantly reduce gas costs by introducing a series of design-level mechanisms, in contrast to code-level gas optimizations~\cite{GASPER17, MadMax18, GASOL20, ibatch21}. These mechanisms ensure efficient app storing and retrieval on the blockchain while minimizing gas consumption. Notably, we achieve this by transitioning from conventional in-contract data structures to transaction log-based contract storage, resulting in a gas reduction by a factor of 53 per operation (20,000 vs. 375 Gas).

\item
Secondly, we surprisingly found that IPFS is not inherently distributed. By default, files are only stored in the original IPFS node and cached at IPFS gateways when there are requests. To achieve true distribution in IPFS, we establish an IPFS consortium network that periodically caches app files at IPFS gateways and performs timely backups of apps at crowdsourced server nodes. Additionally, we propose a mechanism to identify fast and uncensored IPFS gateways for the distributed downloading of apps.
\end{itemize}

To make \name secure and sustainable, we further address three \name-specific system issues.
Firstly, to securely retrieve apps from existing app markets without a network security guarantee, we extract and validate checksums that may be embedded in apps' market pages.
In cases where checksums are unavailable, we implement alternative security measures.
Secondly, to prevent repackaged apps from polluting our market, we propose a lightweight yet effective app certificate field mechanism to detect such apps.
We validate this mechanism experimentally using 15,297 pairs of repackaged apps.
Lastly, to incentivize crowdsourced server nodes, we design a mechanism that charges upload fees, ensuring the self-sustainability of the platform.

We have implemented a prototype of \name using the widely-used Ethereum blockchain~\cite{buterin2014next} and its layer-2 network, Polygon~\cite{PolyLight}.
The implementation comprises 2,485 lines of code written in Solidity, Python, Java, and JavaScript.
In our evaluation, we empirically demonstrate the security effectiveness of \name, successfully preventing man-in-the-middle and repackaging attacks on our app delegation.
We also assess its decentralization by discovering IPFS gateways in over 21 different locations worldwide.
Additionally, we conduct experimental measurements of performance and gas costs.
On average, \name introduces a 12\% performance overhead in tests involving 200 apps from seven app markets.
The cost of app upload is approximately 0.002821 Matic (equivalent to 0.002 USD when one Matic is valued at around 0.7 USD).
We also introduce a batch upload mechanism that reduces gas costs by a factor of 2.02 (for a batch of 10 uploads) to 2.65 (for a batch of 100 uploads).
Notably, \name does \textit{not} require any gas for app downloads as they do not alter the contract state.

%
%
To summarize, this paper presents the following contributions:
\begin{itemize}
\item A novel blockchain-based gateway design (\mysec\ref{sec:design}):
    We introduce a gateway that facilitates trustworthy app delegation, providing permanent, distributed, and secure access to the apps stored in existing markets (and custom apps upload via GitHub URLs).
    Our idea of combining traditional IT infrastructure with decentralized blockchain technology opens a new door for future advancements in blockchain systems.

\item Addressing previously neglected challenges (\mysec\ref{sec:design} and \mysec\ref{sec:sustain}):
    We propose mechanisms to achieve gas-efficient smart contracts and a distributed IPFS design.
    Furthermore, we overcome three specific system issues unique to \name.
  
\item Implementation and extensive evaluation (\mysec\ref{sec:implement} and \mysec\ref{sec:evaluate}):
    We implement a prototype of \name on the Ethereum blockchain and its layer-2 network, Polygon.
    Through extensive evaluation, we assess the performance, gas costs, security, and decentralization of \name, validating its effectiveness in these areas.
\end{itemize}

\textbf{Availability.}
The source code of \name will be released on GitHub after the paper review, facilitating its reuse by future blockchain systems.

\section{Motivation and Background}
\label{sec:motivate}

In this section, we motivate the need of permanent, distributed, and secure app access by measuring the status quo of existing app markets from \mysec\ref{sec:delistedapps} to \mysec\ref{sec:insecureDownload}.
We also provide the necessary background information in \mysec\ref{sec:background} to comprehend our blockchain-based design.

\subsection{Delisted Apps on Google Play}
\label{sec:delistedapps}

It is unclear how many uploaded apps were once delisted.
To estimate this percentage, we conducted a specific measurement on a set of 8,359 popular apps collected from Google Play in November 2018~\cite{DSDK21}.
These apps had one million installs each.
After one year, in November 2019, we re-crawled the apps in the same country and discovered that a significant portion, as high as 13.7\% (1,146 apps), had been delisted during that time period.
This highlights the seriousness of delisted apps in existing app markets and emphasizes the need for a permanent app access mechanism to cater to users who require access to those delisted apps.
\name provides users with the ability to leverage its delegation to permanently store an app on the blockchain before it is delisted.

\subsection{Apps with Limited Global Access}

An easy observation is that many apps on Google Play and Apple Store are limited to specific countries.
For example, the TVB media app~\cite{TVBban} is restricted to Hong Kong, while the popular Hulu app~\cite{Huluban} is available only in the US and Japan.
Additionally, there are limitations in accessing English-based apps in Chinese app markets, and vice versa.
Although these restrictions may be reasonable from the developers' perspective, many users, particularly those traveling or seeking foreign apps, actively search for methods to bypass these limitations.
They resort to online tutorials that involve switching their iTunes accounts to different countries~\cite{APPLEbypass} or using VPNs to circumvent Google Play's restrictions~\cite{PLAYbypass}.
It is important to note that the problem of limited global app access is further compounded by network-based censorship, where certain regions impose restrictions on Google Play~\cite{GooglePlayBan}, denying access to affected users.
While we acknowledge the legitimacy of these various controls, it is essential to develop a mechanism for distributed app access as an alternative solution for users to choose.

\subsection{Insecure App Downloads in Third-Party App Markets}
\label{sec:insecureDownload}

An unexpected observation is that not all app markets provide secure app downloads as Google Play and Apple Store.
Surprisingly, we discovered that half of the top 14 Chinese Android app markets still utilize insecure app downloading via HTTP.
This opens the door for potential injection of repackaged apps~\cite{DroidMOSS12}, particularly when users connect to public Wi-Fi networks~\cite{PublicWiFi} or fall victim to network hijacking~\cite{BGPHijacking13}.
This poses a severe security risk, especially considering that around one billion Internet users in China rely on third-party app markets due to the ban on Google Play.

\begin{table}[t!]
\caption{The measurement result of app downloading security in the top 16 Chinese Android app markets~\cite{ChineseAppMarkets18}.}
\label{tab:3rdmarkets}
\begin{adjustbox}{center} 
\scalebox{0.85}{
\begin{tabular}{ccc}

 \toprule
  Market Name  & Company Type & Secure App Downloading?\\ 

 \midrule
  Tencent Myapp & Web Co. & HTTPS \gou \\
  Baidu Market & Web Co. & HTTP \cha \\
  360 Market & Web Co. & HTTP \cha \\

 \midrule
  OPPO Market & HW Vendor & No Website Download\\
  Huawei Market & HW Vendor & HTTPS \gou \\
  Xiaomi Market & HW Vendor & HTTPS \gou \\
  Meizu Market & HW Vendor & HTTP \cha \\
  Lenovo MM & HW Vendor & HTTP \cha \\

 \midrule
  HiApk & Specialized & Cannot Find the Website \\
  Wandoujia & Specialized & HTTPS \gou \\
  PC Online & Specialized & HTTPS \gou \\
  LIQU & Specialized & HTTPS \gou \\
  25PP & Specialized & HTTPS \gou \\
  App China & Specialized & HTTP \cha \\
  Sougou & Specialized & HTTP \cha \\
  AnZhi & Specialized & HTTP \cha \\
 \bottomrule

\end{tabular}
}
\end{adjustbox}
\end{table}

Table~\ref{tab:3rdmarkets} presents our measurement results for the top 16 Chinese Android app markets, as identified by a recent study on app markets~\cite{ChineseAppMarkets18}.
Specifically, we examined whether these markets employ HTTPS as the medium for hosting app downloads.
It is important to note that while some market websites may use HTTPS for hosting web content, they may still provide app downloads exclusively through insecure HTTP.
To distinguish this situation, we focused solely on app downloading traffic.
As indicated in Table~\ref{tab:3rdmarkets}, half of the 14 Chinese app markets (excluding two markets that do not offer web-based app downloading) utilize insecure HTTP downloading.
These markets include prominent Internet giants (e.g., Baidu) and smartphone hardware vendors (e.g., Meizu and Lenovo), as well as specialized app markets (e.g., Anzhi and App China).


\subsection{Relevant Technical Background}
\label{sec:background}

To facilitate permanent, distributed, and secure app access, we propose a novel architecture based on Ethereum (Polygon) and IPFS. In this subsection, we provide the necessary background information to better understand our proposed solution.

\textbf{Blockchain.}
A blockchain is typically a public and distributed ledger. 
It records transactions that are immutable, verifiable, and permanent~\cite{Blockchain16}. 
Therefore, blockchain can be utilized as a decentralized database. 
The trust among different nodes is guaranteed by the so-called \textit{consensus} (e.g., Proof of Work) instead of the authority of a specific institution. 
Consensus is the key for all nodes on a blockchain to maintain the same ledger in a way that the authenticity could be recognized by each node in the network.

\textbf{Ethereum}~\cite{buterin2014next} is the second largest blockchain system and the most popular smart contract platform~\cite{MineBlkVuln22, BlockScope23}.
A smart contract is a contract that has been programmed in advance with a sequence of rules and regulations for self-executing.
Solidity is the primary language for programming smart contracts on Ethereum.
In particular, it is a Turing-complete language, which suggests that developers could achieve arbitrary functionality on smart contracts theoretically. 
To prevent denial-of-service attacks, users need to consume gas fees to send transactions on Ethereum.
The gas fees are paid in Ethereum's native cryptocurrency called Ether (or ETH).
Recently, \textbf{Polygon}~\cite{PolyLight}, Ethereum's layer-2 network, has been emerging as a result of the expensive gas fee and low throughput in Ethereum.

\textbf{IPFS} (Interplanetary File System)~\cite{IPFS14} is a peer-to-peer file sharing system, where files are stored in a distributed way and routed using the content-addressing \cite{curran2018interplanetary}. 
IPFS was proposed because the storage of large files on blockchain is inefficient and with high costs. 
Specifically, all nodes in a blockchain network need to endorse the entire ledger and synchronize the files stored. 
As a result, unnecessary redundancy will lead to a huge waste of storage, and the latency of ledger synchronicity will also be significantly increased. 
To address this limitation, we can store data only in the IPFS storage nodes and keep the unique and permanent IPFS address (called IPFS hash) in the blockchain.

\section{The Core Design of AGChain}
\label{sec:design}

In this section, we present the core design of \name, including its objectives, threat model, overall design, and the two key challenges we tackled.

\subsection{Design Objectives and Threat Model}
\label{sec:objective}

\textbf{Design objectives.}
Our goal is to develop a trustworthy blockchain-based gateway that enables permanent, distributed, and secure app delegation from existing app markets. It is worth noting that our intention is not to replace these existing markets, as the apps on \name ultimately originate from them. Instead, our aim is to provide an additional option for users with security concerns or those who wish to back up their apps. They can utilize \name as a secure proxy to download apps from third-party markets or store them permanently on the blockchain. Therefore, \name functions more as a gateway to existing app markets rather than a standalone market. To achieve this, we have identified the following key design objectives:

\begin{itemize}
  \item \textit{Permanent delegation.}
To achieve the permanent storage and delegation of apps on \name, we utilize the immutability of blockchain technology. In doing so, we make two specific choices. Firstly, instead of developing our own blockchain infrastructure like Infnote~\cite{Infnote20}, we leverage existing and well-established blockchains such as Bitcoin and Ethereum. These mainstream blockchains have a large network of nodes, accumulated over time, making them robust against various attacks. Secondly, to implement our logic on the blockchain, we opt for writing a smart contract rather than creating a virtualchain as seen in ~\cite{IoTDataBlockchain17, Blockstack16, Virtualchain16}. Smart contracts are lightweight yet powerful in terms of their Turing-completeness. Note that only apps accessed through \name can benefit from permanent access.

  \item \textit{Distributed delegation.}
Due to the high storage and gas costs associated with directly storing raw files on the blockchain (as discussed in \mysec\ref{sec:background}), we need an efficient and distributed file storage solution. In this paper, we utilize IPFS for its distributed and permanent nature, unlike centralized cloud services that cannot guarantee continuous availability (some are even subject to censorship, such as Google Drive and Dropbox). The basic idea behind incorporating IPFS into \name is to store raw app files on IPFS while keeping their corresponding IPFS indexes on the blockchain. However, we discovered an unexpected behavior of IPFS, where files are not duplicated to other IPFS nodes unless there is a request for them. To truly achieve distribution, we establish a consortium network by leveraging IPFS gateways and crowdsourced server nodes.

  \item \textit{Secure delegation.}
A practically desirable objective is to enable secure app download delegation for apps obtained from markets that do not support HTTPS downloading (referred to as \textit{unprotected markets} hereafter). This objective has significant implications for millions of Chinese market users (as discussed in \mysec\ref{sec:insecureDownload}). Specifically, we aim to securely retrieve apps from existing app markets without relying on a dedicated and trusted network path. In addition to ensuring download security, we also need to prevent repackaged apps~\cite{DroidMOSS12} being uploaded to \name.

\end{itemize}

Besides the core objectives, we also aim for \textit{sustainable delegation} to ensure \name's self-sustainability by providing monetary incentives to crowdsourced server nodes. The implementation details of secure delegation and sustainable delegation will be explained in \mysec\ref{sec:sustain}. In this section, our focus is on achieving permanent and distributed delegation for \name.

\textbf{Adversary model.}
In this paper, we assume adversaries:

\begin{itemize}
  \item \textit{Cannot break TLS (Transport Layer Security) and hash (e.g., SHA-256) algorithms.}

  \item \textit{Cannot compromise the underlying (Ethereum and Polygon) blockchain and IPFS storage.}

  \item \textit{Cannot exploit our smart contract and the core server code.}
    We can leverage the recent advances on verification of smart contracts~\cite{Securify18} and write SGX-enabled\footnote{SGX (Software Guard eXtension) is a TEE (trusted execution environment) technique developed by Intel to protect selected code and data.} server code~\cite{RUSTSGX19} to enhance our contract and server code's security, although this is out of the scope of this paper.
\end{itemize}

We do, however, assume that attackers \textit{can} intercept unprotected network traffic, upload repackaged apps to original app markets, compromise or replace our front-end code (since it is at the user side), access our smart contract, etc.

\subsection{The Overall System Design}
\label{sec:overall}

\begin{figure*}[t!]
\begin{adjustbox}{center}
\includegraphics[width=\textwidth]{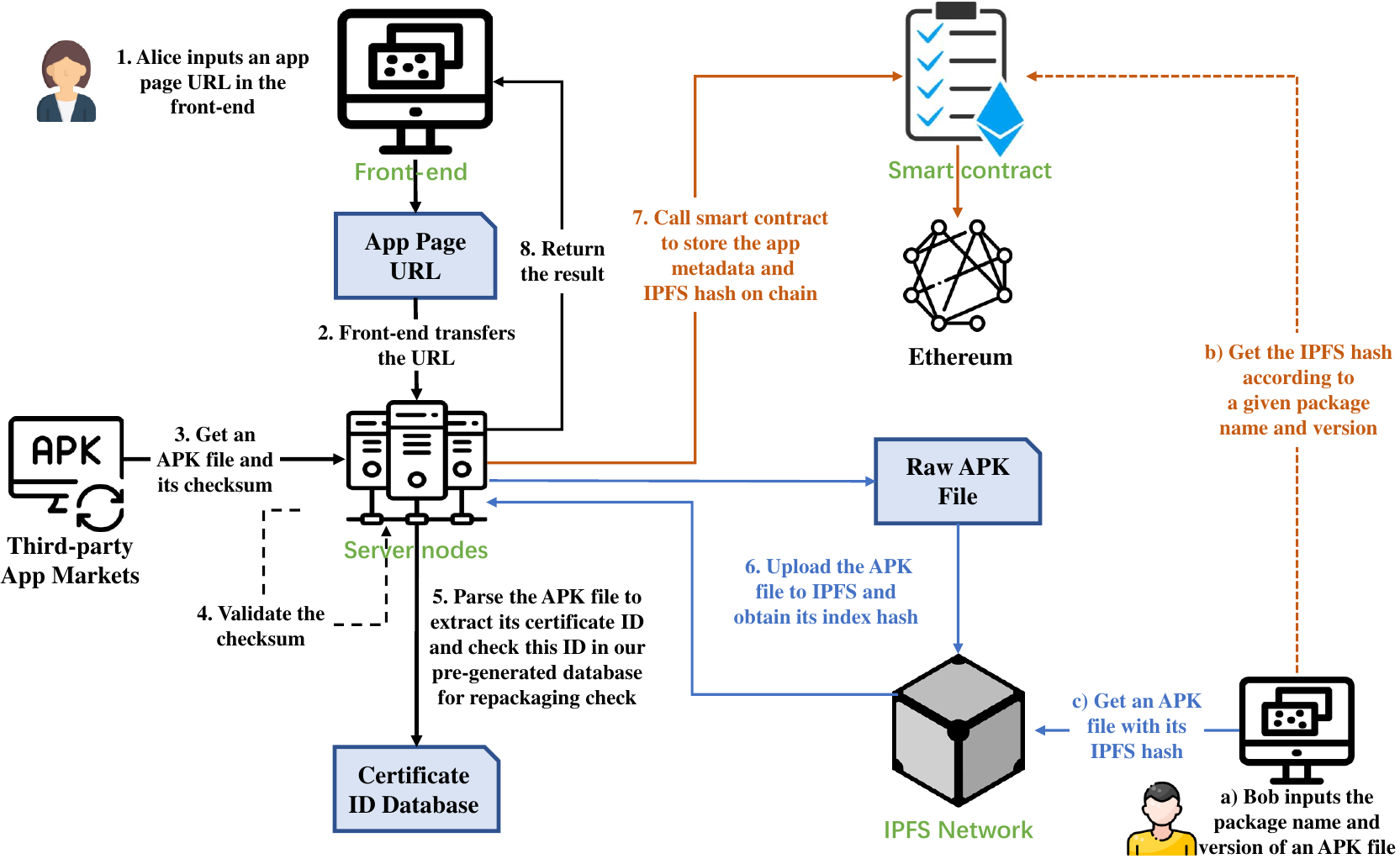}
\end{adjustbox}
\caption{A high-level design and the workflow of \name, which consists of four components marked in the green color.}
\label{fig:overall}
\end{figure*}

\textbf{Architecture.}
Figure~\ref{fig:overall} presents \name's high-level design.
As highlighted in the green color, it has four components as follows:

- \textit{Smart contract.}
The most important component is a novel (gas-efficient) smart contract, which stores all app metadata on chain and duplicates them on most of the Ethereum nodes worldwide.
With these data (including IPFS file indexes) and their programmed storing and retrieving logic, this smart contract is the actual control party of \name's entire logic.

- \textit{IPFS network.}
Another core component is an IPFS (consortium) network, which stores raw app files in a distributed manner.
The stored apps then can be automatically routed and retrieved through IPFS's content-addressing~\cite{IPFSContentAddressing}.
Note that with the smart contract and IPFS components, we guarantee the permanent and decentralized app access in \name.

- \textit{Server nodes}.
To achieve app download security and repackaging checking, we also need server(s) to retrieve apps from existing markets, inspect their security, and schedule their uploading as these tasks cannot be performed in blockchain.
Since any machine with our server code could be a server node, we propose an incentive mechanism (details in \mysec\ref{sec:chargeFees}) to motivate servers to join \name.
These \textit{crowdsourced} server nodes further enhance the decentralization.

- \textit{Front-end.} Finally, we provide a front-end web interface to help uploaders and downloaders interact with \name. Note that for app downloading, our front-end directly communicates with smart contract and IPFS without the server.

\textbf{Upload workflow.}
Figure~\ref{fig:overall} also shows the overall workflow of \name.
As shown in left part, Alice wants to securely download an app (e.g., Alice is a security-sensitive user) or permanently store an app on chain for future usage (e.g., Alice is a developer or a user who wants to backup the current version of an app). 
She then acts as an app uploader:

\begin{itemize}
	\item[1.] Alice just needs to input the original app page URL and clicks the ``upload'' button in the front-end. \name automatically finishes all the remaining steps.

	\item[2.] The front-end transfers the URL to one server node.

	\item[3.] After knowing the URL, the server analyzes the corresponding app page to obtain the download URL that points to the actual APK file (the file format used by Android apps), and retrieves it from its original market.

    \item[4.] For the app markets using insecure app downloading (e.g., those seven markets in Table~\ref{tab:3rdmarkets}), \name performs one more step to check whether the retrieved file has been tampered with during network transmission.

    \item[5.] For all third-party markets, we conduct repackaging checks to prevent repackaged apps~\cite{DroidMOSS12} from polluting \name.
      We propose a lightweight yet effective certificate ID based mechanism; see details in \mysec\ref{sec:repackageDetection}.
      During this process, we also parse the APK file to extract its package name and version (besides the certificate ID).

	\item[6.] The server then uploads the raw APK file to IPFS and obtains its corresponding IPFS hash (i.e., the file index).

	\item[7.] Finally, the server invokes the smart contract to store all app metadata and IPFS hash on chain. We choose the widely-used Ethereum and its layer-2 network Polygon (see \mysec\ref{sec:background}) as our underlying blockchain.

    \item[8.] To avoid Alice from waiting a long time for blockchain transaction confirmation (typically 9$\sim$13 seconds, according to our tests), \name simultaneously returns the result of package name, app version, and IPFS hash.
\end{itemize}

\textbf{Download workflow.}
As shown in the right part of Figure~\ref{fig:overall}, Bob acts as an app downloader to download apps (e.g., the app uploaded by Alice) that are already in \name:

\begin{itemize}
\item[a)] Bob inputs (or browses) the package name and version of the app that he wants to download in the front-end.

\item[b)] The front-end then automatically invokes the smart contract to retrieve the corresponding IPFS hash.

\item[c)] The front-end further locates the nearest IPFS network node to download the APK file and returns it to Bob.
\end{itemize}

\textbf{Major challenges.}
In the course of developing \name, we identify two previously neglected challenges:

\begin{description}
  \item[C1:] \textit{How to minimize gas costs in the smart contract?}
As each app upload requires a blockchain transaction and incurs gas fees, it is crucial to minimize these costs in our smart contract. Although there have been some code-level gas optimizations~\cite{GASPER17, MadMax18, GASOL20, ibatch21}, they are still insufficient. In \mysec\ref{sec:contractdesign}, we introduce design-level optimizations that significantly reduce gas costs by a factor of 16.

  \item[C2:] \textit{How to enable fully distributed IPFS storage?}
As discussed in \mysec\ref{sec:objective}, we discovered that IPFS files are only stored in the original IPFS node and cached at IPFS gateways upon request. If the original node goes offline, the file becomes inaccessible to the entire IPFS network (availability is restored when another node adds the same content). To achieve distributed IPFS storage, we establish an IPFS consortium network (\mysec\ref{sec:ipfsdesign}) that proactively requests IPFS gateways and crowdsourced server nodes to back up files.
\end{description}


\subsection{Gas-Efficient Smart Contract for App Metadata Storing and Retrieving}
\label{sec:contractdesign}

In this subsection, we introduce a series of mechanisms aimed at reducing gas costs for app metadata storage (i.e., app uploads), eliminating gas costs for metadata retrieval (i.e., app downloads), and implementing a whitelist mechanism to prevent misuse of our contract functions. These design-level optimizations are significantly more efficient than code-level optimizations~\cite{GASPER17, MadMax18, GASOL20, ibatch21} and provide valuable guidance for future smart contract designs.



\textbf{Minimizing gas costs via log-based contract storage.}
We found that a major source of gas inefficiency comes from data storage in the smart contract.
Like many other smart contracts, \name needs to store app metadata as a structure defined in the contract.
However, such data storage operation, via the \texttt{SSTORE} instruction in Ethereum Virtual Machine (EVM), changes the internal block states and Ethereum's world state~\cite{WorldState}.
Therefore, it is expensive, costing 20,000 Gas\footnote{The gas cost/fee is the product of gas price (or Gwei) and the Gas consumed.} per operation~\cite{EthereumYellow14}. 
Since numerous app metadata records will be stored in \name, it would cost a large amount of gas fees if we use the traditional contract structure. 
Fortunately, we identify a logging interface in the EVM, which can be used to permanently log data in transaction receipts via the \texttt{LOG} instruction~\cite{EthereumYellow14}.
Since it changes only the block headers and does not need to change the world state, only 375 Gas is consumed per operation.
The underlying logging mechanism is complicated, and we refer interested readers to the Ethereum yellow paper~\cite{EthereumYellow14, EthLog} for more details.
While log-based contract storage dramatically reduces gas fees, it has no structure information.
We thus ask our server nodes to recover the app metadata structure, which includes the app package name, app package version, app certificate serial number, the original market page URL, the repackaging detection result, and the important IPFS hash.

\textbf{Offloading on-chain duplicate check to server nodes.}
Another major source of gas costs originates from the duplicate check, which checks duplicated app metadata before storing it on chain.
Originally, we deployed such a check in the smart contract, but we found that it is costly since each check needs to iterate over the entire app metadata structure.
Moreover, an in-contract structure has to be maintained, which causes the first log-based optimization unadoptable. 
Therefore, we offload this on-chain duplicate check to server nodes, which query the latest app metadata from the smart contract before uploading any records.
If a duplicate exists, the uploading will stop.


\textbf{Further reducing gas costs via batch uploads.}
With the above two default optimization mechanisms, we significantly reduce gas costs by a factor of 15.75 (0.001347 v.s. 0.0000855 Ether, before and after the optimization).
We further reduce the costs by providing a back-end interface of batch uploads.
Our experiment shows that by batching 10 uploads together, we save gas by a factor of 2.02 (compared with 10 times of normal uploads).
This factor increases to 2.65 for a batch upload of 100 records. 
These suggest that for a large number of app uploads (e.g., a company uploads all its apps), we can use batch uploads to minimize gas costs.

\textbf{Eliminating gas costs for all app downloads.}
While app uploading certainly consumes gas, we find a way to eliminate gas costs for all app downloads.
For a normal contract data structure, we originally used the \texttt{view} function modifier to describe the data retrieving contract function since it does not change any contract state. 
Invoking such a view-only function (even by external parties) will not initiate any blockchain transaction, and thus no gas fee is needed.
However, since we have switched to log-based contract storage, we create a bloom filter~\cite{EthBloom, EthLog} to quickly locate the block headers containing our data logs, regenerate the original logs, and extract IPFS hashes from them.
Since this task is performed only at the server side via the web3 Python APIs, the contract side will not cost any gas. 

\textbf{Achieving a whitelist mechanism for access control.}
Since smart contract has no access control mechanism, a contract function can be invoked by anyone.
This implies that an adversary can invoke our \texttt{storeApp()} function to upload any app in our scenario.
To present malicious uploads from interfering \name's data, we implement a lightweight whitelist mechanism that consists of two function modifiers, \texttt{onlyOwner(address caller)} and \texttt{onlyWhitelist(address caller)}, where the parameter \texttt{address} is the invoking party's Ethereum account address.
By enforcing the \path{onlyWhitelist} \path{(msg.sender)} modifier to check the caller of \texttt{storeApp()}, we can guarantee that only an account in the whitelist can upload apps.
We also implement two contract functions to add or delete an account address from the whitelist, and they are enforced by the \texttt{onlyOwner(msg.sender)} modifier.
Note that the owner is our contract creator, and we gradually add each authorized server node into the whitelist.
By designing such a hierarchical whitelist, we can achieve effective access control to avoid malicious data injecting into \name.

\subsection{IPFS Consortium Network for Distributed File Uploading and Downloading}
\label{sec:ipfsdesign}


To overcome challenge C2, \name utilizes IPFS gateways and crowdsourced server nodes to cache or backup apps, ensuring their true distribution within the IPFS network. By establishing these gateways and servers, \name effectively creates an IPFS consortium network, enabling distributed app access.

\textbf{Periodically caching app files at IPFS gateways.}
As described in challenge C2, an IPFS file is only accessible through the original IPFS node and is distributed via content-addressing~\cite{IPFSContentAddressing}. However, if the original node becomes offline, the file becomes inaccessible to the entire IPFS network. Fortunately, our experiment revealed that an IPFS gateway caches files for a certain duration when accessed through the gateway. Leveraging this observation, we intentionally simulate user requests to cache apps at IPFS gateways. To achieve this, we deploy a script on the server that sends periodic file requests through various IPFS gateways. These requests are sent before the IPFS garbage collector cleans our app files, ensuring that copies of the app files are always available in IPFS gateways.

\textbf{Timely backing up apps at crowdsourced server nodes.}
To achieve fully distributed app storage, we utilize each crowdsourced server as an IPFS storage node and ensure timely backups of apps in our IPFS consortium network. Each server node initializes by running the \texttt{ipfs daemon} command to function as an IPFS node. It then executes a consortium synchronization script, which obtains a list of IPFS hashes for the apps in \name and retrieves the corresponding raw app files from the IPFS network. To prevent the IPFS garbage collection from removing these files, the script locally pins the raw apps using the \texttt{ipfs pin} command. This approach enhances data redundancy in \name and enhances the distribution of app storage.

\textbf{Identifying fast IPFS gateways for app downloading.}
In addition to distributed app uploads, we introduce a mechanism to enable distributed and fast app downloading in \name. The front-end of \name performs RTT (Round-Trip Time) tests on public IPFS gateways and selects the gateway with the lowest RTT for downloading the raw app file. This approach enhances the performance of app downloading in IPFS while also mitigating the risk of potential censorship by certain IPFS gateways.

%
%
%

\section{Making It Secure and Sustainable}
\label{sec:sustain}

So far, we have designed \name to be permanent and distributed through contract-based distribution and IPFS-based storage.
However, to make it secure and sustainable, we still need to address three \name-specific system issues: 

\begin{description}
  \item[C3:] \textit{How to securely retrieve apps from the app markets without a network security guarantee?}
    Recall that our server needs to retrieve apps from existing markets before uploading them to IPFS.
    This process is relatively simple for markets that use HTTPS, but it becomes challenging for those that rely on insecure HTTP (as discussed in \mysec\ref{sec:insecureDownload}) since there is no inherent network security guarantee.
    To tackle this issue, we propose two modes of secure app retrieval in \mysec\ref{sec:secureRetrieval}.

  \item[C4:] \textit{How to avoid repackaged apps from polluting our market?}
    By addressing challenge C3, we ensure that the retrieved app is identical to the one available in its original app market. 
    However, in the case of a third-party app market, there is a possibility that the app has already been repackaged before our secure retrieval. 
    Consequently, we require a mechanism, as proposed in \mysec\ref{sec:repackageDetection}, to detect repackaged apps~\cite{DroidMOSS12} that may differ from their official versions on Google Play.

  \item[C5:] \textit{How to make \name self-sustainable?}
    As stated in \mysec~\ref{sec:overall}, each server node in the crowdsourced network consumes computer resources and incurs gas fees for app uploading in \name. 
    This becomes unsustainable without a monetary mechanism to incentivize these nodes. 
    Considering that uploaders benefit from \name by advertising their apps or securely storing them for security research purposes, it is reasonable to impose an upload fee on them to compensate the crowdsourced servers. 
    We will introduce this mechanism in \mysec\ref{sec:chargeFees}. 
    By implementing upload fees, we can also deter spammers from abusing \name.
\end{description}

\subsection{Secure App Retrieval from (Existing) Unprotected Markets}
\label{sec:secureRetrieval}

In this subsection, we present two methods that collectively achieve secure app retrieval even for those unprotected markets, e.g., seven markets using HTTP downloading in Table~\ref{tab:3rdmarkets}.

\textbf{Secure app downloading via checksums.}
We find that although those seven markets use HTTP to download apps, most of them allow users to browse app pages via HTTPS (e.g., \url{https://zhushou.360.cn/detail/index/soft\_id/95487}).
Additionally, we can extract app APK file checksums (e.g., MD5) from the HTML source code of these pages. 
For instance, three app markets (Baidu, 360, and AppChina) directly embed the checksums in their app download URLs. 
In the case of Lenovo MM, the checksum is embedded in the \texttt{<script>} data section of the app's HTML page.
By analyzing the HTML pages of these markets, we can securely obtain app checksums and compare them with the calculated checksums of the app APK files we retrieve via HTTP. 
If they match, we conclude that the retrieved app has not been tampered with by adversaries. 
Furthermore, although Sogou market does not provide checksum information, we discover that its HTTP download URL can be converted into an HTTPS version. 
Consequently, we can ensure guaranteed app download security for five out of the total seven unprotected markets.

\textbf{Alternative security with no checksums.}
To address the lack of checksums in the remaining app markets such as Meizu and Anzhi, we propose an alternative mechanism to ensure download security. 
The fundamental concept is to cross-check each downloaded app with its Google Play counterpart. 
This can be achieved by utilizing the AndroZoo app repository~\cite{AndroZoo16}, which houses a vast collection of over ten million apps sourced from Google Play.
By comparing a downloaded app with the apps in the AndroZoo repository, we can determine whether it is present. 
If it is, we conclude that the downloaded app from a third-party market has not been compromised during the download process. 
Additionally, even if an app is not found in the repository, we can extract its developer certificate and verify if it belongs to a Google Play developer.
Given the dynamic nature of the AndroZoo repository, which continuously updates to reflect changes in Google Play, we have a high level of confidence in covering most apps through either app-level or developer-level checking. 
For the rare cases where some apps are not adequately covered, \name will issue a warning indicating that these apps might not have been securely retrieved.

\subsection{Exploiting App Certificate Info for Repackaging Detection}
\label{sec:repackageDetection}

In this subsection, we present a mechanism of exploiting app certificate to \textit{accurately} detect repackaged apps that might be uploaded to \name from third-party app markets.

In contrast to traditional app repackaging detection methods~\cite{DroidMOSS12, Repack19, WardenAttack21}, the identifier of apps uploaded to \name, namely the app package name, remains fixed. 
This fixed package name identifier allows us to locate the corresponding official app on Google Play. 
The challenge lies in differentiating between these two versions of apps and specifically determining whether they originate from the same developer.
To address this challenge, we examine the data structure of the app certificate, which includes essential fields such as the certificate serial number, issuer, subject, and X509v3 extensions. 
Among these fields, we identify the serial number (e.g., 0x706a633e) as the most lightweight metric that adversaries cannot manipulate since they lack access to the developers' app signing key. 
On the other hand, fields like issuer and subject are susceptible to manipulation, and X509v3 is more complex compared to the serial number.

We further experimentally validate our detection idea by utilizing a dataset~\cite{Repack19} consisting of 15,297 \textit{pairs} of repackaged Android apps. 
Each pair consists of an original app and its corresponding repackaged version. 
This dataset serves as our ground truth. 
To facilitate our analysis, we developed a script that automatically extracts the package name and serial number from any given app APK file. 
This script utilizes the widely used Androguard library for parsing APK files. 
It is worth noting that we have integrated this script into \name's server code.
By running our script on the 15,297 app pairs, we discover a total of 2,270 pairs that share the same package name, while none of the pairs have the same serial number. 
Moreover, for the remaining 13,027 repackaged pairs, each pair has distinct package names. 
This experiment provides compelling evidence that our serial number-based mechanism achieves 100\% accuracy in detecting repackaged apps.


\subsection{Charging Upload Fees to Maintain the Platform Self-Sustainability}
\label{sec:chargeFees}

In this subsection, we design a mechanism of charging app upload fees to pay crowdsourced server nodes in \name so that the entire \name platform is sustainable.

We thus revise the original design of \name to charge upload fees just before each server node invokes (IPFS and) smart contract.
More specifically, we add two more steps between step 5 and 6 in \myfig~\ref{fig:overall}.
The first step is to send an estimated transaction fee to our smart contract, for which we add one more function called \texttt{DonateGasFee()} in our smart contract.
We set this function \texttt{payable} so that the user side (in the form of our JavaScript code) can invoke it with a fee, e.g., \texttt{DonateGasFee().send(from: userAccount, value: fee)}.
After this call is successfully executed in blockchain, the user side will receive a transaction ID.
In the second step, our JavaScript code automatically sends this transaction ID (and a few other metadata) to the server for verification.

We now explain how our JavaScript code estimates the gas fee and how the server verifies the payment transaction.
To calculate a gas fee at the user side, we invoke a web3 JavaScript function called \texttt{estimateGas()} to estimate the required gas of executing the transaction in the EVM.
For the transaction verification at the server, we first query the transaction according to its ID, and then determine (i) whether the destination address of this transaction is our smart contract, (ii) whether the upload fee exceeds our estimated gas, and (iii) whether this transaction is never used before.
Only when the three conditions are all satisfied, the server then continues the actual app uploading to IPFS and Ethereum/Polygon.

\section{Implementation}
\label{sec:implement}

\begin{figure}[t!]
\begin{adjustbox}{center}
\includegraphics[width=0.6\textwidth]{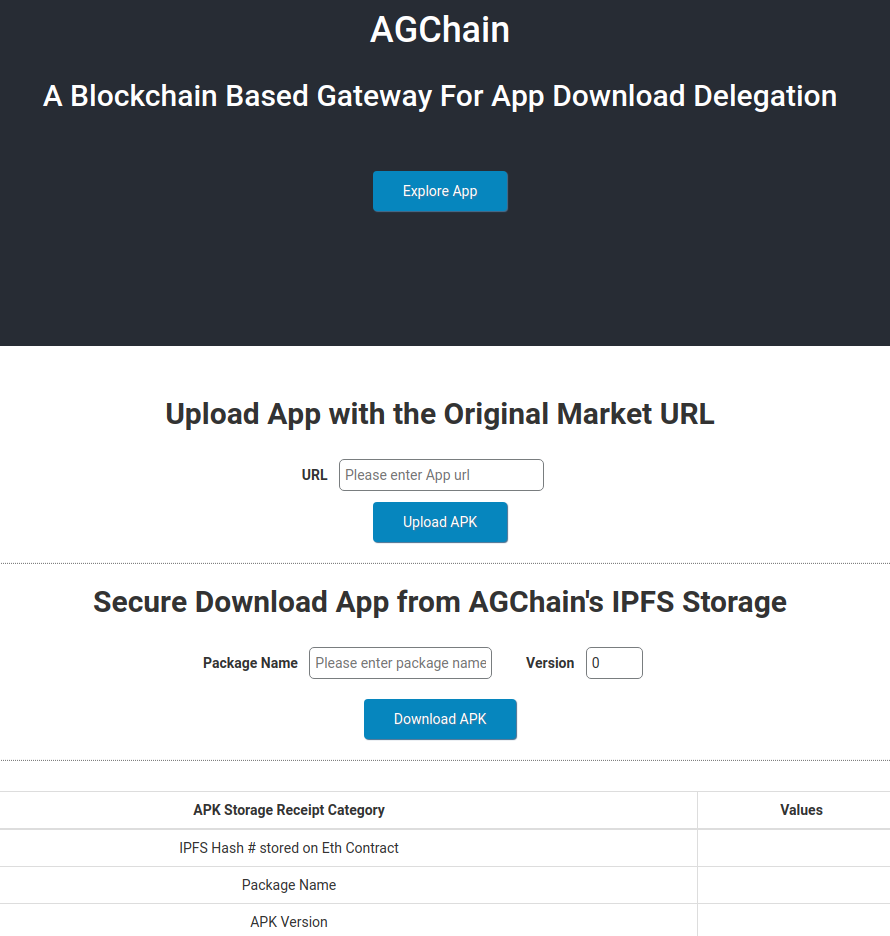}
\end{adjustbox}
\caption{A screenshot of \name's front-end homepage (cutted). Besides uploading and downloading apps, users can click the ``Explore App'' button to browse the apps stored in \name. The bottom part will output the metadata of each new upload, which can help users immediately download the app from \name using the ``Download APK'' button.}
\label{fig:homepage}
\end{figure}

We have implemented a prototype of \name, utilizing Ethereum smart contract and IPFS. 
\myfig~\ref{fig:homepage} shows a screenshot of the front-end homepage.
In this section, we summarize the implementation details of \name.
The current prototype consists of 2,485 LOC (lines of code), excluding all the library code used.
Table~\ref{tab:LineOfCode} lists a breakdown of LOC across different components and programming languages.

\begin{table}[t!]
\caption{A breakdown of LOC (lines of code) in \name.}
\label{tab:LineOfCode}
\begin{adjustbox}{center}
\scalebox{0.85}{
\begin{threeparttable}
\begin{tabular}{|c|c|c|c|c|c|}

\cline{2-6}
\multicolumn{1}{c|}{}   & Front-end & Server    & Contract  & IPFS  & \multirow{2}{*}{\name} \\
\multicolumn{1}{c|}{}   & (F)       & (S)       & (C)       & (I)   & \\
\hline

JavaScript              & 781       &           & (302 in F)$^*$& (80 in F) & 781 \\
\hline

Java                    &           & 849       &           & (36 in S) & 849 \\
\hline

Python                  &           & 723       & (445 in S)&       & 723 \\
\hline

Solidity                &           &           & 51       &       & 51 \\
\hline

CSS                     & 81        &           &           &       & 81 \\
\hline

Sum                     & 862       & 1,572     & 51 + (747) & (116)& 2,485 \\
\hline

\end{tabular}
\begin{tablenotes}
\item $^*$This means that 302 lines of code in front-end are related to smart contract. \hspace*{1.1ex}Other brackets, such as (445 in S) and (80 in F), are similar.
\end{tablenotes}
\end{threeparttable}
}
\end{adjustbox}
\end{table}

%
%
%

\textbf{Front-end.}
We implement the front-end user interface using the React web framework. 
Hence, the HTML and CSS code is minimal, and some HTML contents are also dynamically generated using JavaScript.
Besides user interfaces, we write 302 JavaScript LOC on top of the web3.js library to query our smart contract for retrieving IPFS hashes. 
To execute IPFS commands in JavaScript for app downloading, we write additional 80 JavaScript LOC based on the js-ipfs library. 
Overall, the front-end implementation consists of approximately 862 lines of code.


\textbf{Server node.}
We implement our server code in a total of 1,572 LOC and run it on the AWS (Amazon Web Services).
Specifically, we write 849 Java LOC to handle requests from the front-end, securely download apps from existing markets, perform repackaging checks, and upload APK files to IPFS.
Moreover, we write 445 Python LOC to leverage web3 Python APIs to interact with smart contract.
Lastly, for APK file parsing (used in repackaging checks), we leverage the Androguard library~\cite{Androguard} and write 278 Python LOC on top of it.

One of the main tasks of the server node is to retrieve apps from the market URLs provided by users. 
However, these URLs often only consist of page URLs, such as \url{https://shouji.baidu.com/software/11569169.html}, instead of direct download URLs. 
In order to automatically extract the download URLs, we utilize the understanding that each market follows a distinct pattern when transitioning from a page URL to a download URL. 
As a result, we conduct pre-analysis of these markets to obtain their respective download URL patterns. 
For the majority of markets, their download URLs can be directly extracted from the HTML tags, similar to the aforementioned Baidu market example. 
However, a few markets, such as Meizu and AnZhi, require the calculation of URLs from their JavaScript code. 
Currently, our prototype has analyzed the download patterns of all seven markets that require \name's secure app download delegation.

\textbf{Smart contract.}
We implement the smart contract with 51 LOC in the Solidity language, which was reduced from 128 LOC in the earlier version since we no longer define in-contract data structures (see \mysec\ref{sec:contractdesign}).
Therefore, our smart contract mainly describes a list of functions, e.g., \texttt{storeApp()} for uploading app metadata to the blockchain.
In particular, to avoid the smart contract being misused by unauthorized parties, we set that only whitelisted server nodes can invoke the \texttt{storeApp()} function by adding a function modifier to check the transaction sender.
However, this also prevents the front-end's \texttt{estimateGas()} web3 API from estimating gas (see \mysec\ref{sec:chargeFees}).
To address this issue, we create an additional function called \texttt{storeApp\_estimate()} that duplicates \texttt{storeApp()}'s functionality but does not execute data push operations. 
This function will be executed by the EVM instead of \name transactions.

\textbf{IPFS module.}
As we do not make any modifications to the IPFS network, the implementation of IPFS-related code is carried out in other components. 
For instance, we have developed 80 JavaScript LOC for the front-end to facilitate the downloading of apps from IPFS. 
Similarly, the server node is responsible for uploading apps to IPFS, and we have implemented this functionality using 36 lines of Java code (excluding file operation code). 
To activate the IPFS node on the server, we simply run the \texttt{ipfs daemon} command.

\section{Evaluation}
\label{sec:evaluate}

In this section, we first experimentally evaluate the performance and gas costs of \name, then empirically demonstrate its security effectiveness and decentralization.
We have deployed \name to the Ethereum blockchain (tested in the Ethereum Rinkeby environment) in 2021 and further deployed it to Polygon (a widely-used Ethereum layer-2 network; see \mysec\ref{sec:background}) in 2022 for a real-world use.

\subsection{Performance}
\label{sec:performance}

				
				
				

To fairly evaluate the additional time introduced by \name, we use our server code to record both normal downloading time (part of step 3 in \myfig~\ref{fig:overall}) and \name's processing time (the rest of step 3 and steps 4 to 6).
Note that we do not count step 8 as part of \name's overhead, because we simultaneously return results to users without waiting for the transaction to be confirmed.
Totally, we conduct 200 app tests from seven markets (one market, Sogou, is no longer available in 2022 when we test in Polygon) that require secure delegation.
Moreover, we perform these tests in different days to minimize the impact of different network conditions.

\begin{table*}[t!]
\caption{Average processing time introduced by \name.}
\label{tab:performance}
\begin{adjustbox}{center} 
\scalebox{0.6}{
\begin{threeparttable}
\begin{tabular}{|c|c|c|c|cccccccc|}
\hline
\multirow{3}{*}{\textbf{Market ID}} &
  \multirow{3}{*}{\textbf{Market Name}} &
  \textbf{APK} &
  \textbf{Normal} &
  \multicolumn{8}{c|}{\textbf{Additional Processing Time (ms)}} \\ \cline{5-12} 
 &
   &
  \textbf{Size} &
  \textbf{Download} &
  \multicolumn{2}{c|}{\textbf{Checksum}} &
  \multicolumn{2}{c|}{\textbf{Repackaging}} &
  \multicolumn{2}{c|}{\textbf{IPFS Upload}} &
  \multicolumn{2}{c|}{\textbf{Overall \%}} \\ \cline{5-12} 
 &
   &
  \textbf{(Mb)} &
  \textbf{Time (ms)} &
  \textbf{E*} &
  \multicolumn{1}{c|}{\textbf{P*}} &
  \textbf{E} &
  \multicolumn{1}{c|}{\textbf{P}} &
  \textbf{E} &
  \multicolumn{1}{c|}{\textbf{P}} &
  \textbf{E} &
  \textbf{P} \\ \hline
1 &
  Baidu Market &
  20.09 &
  71063.2 &
  820.1 &
  \multicolumn{1}{c|}{1324.9} &
  377.1 &
  \multicolumn{1}{c|}{557.8} &
  340.7 &
  \multicolumn{1}{c|}{599.5} &
  2.16\% &
  3.49\% \\
2 &
  360 Market &
  27.26 &
  10105.5 &
  917.0 &
  \multicolumn{1}{c|}{1333.8} &
  433.8 &
  \multicolumn{1}{c|}{393.0} &
  429.1 &
  \multicolumn{1}{c|}{351.2} &
  17.61\% &
  20.56\% \\
3 &
  Lenovo MM &
  20.87 &
  14819.2 &
  1203.3 &
  \multicolumn{1}{c|}{1245.3} &
  408.6 &
  \multicolumn{1}{c|}{391.9} &
  415.8 &
  \multicolumn{1}{c|}{277.5} &
  13.68\% &
  12.92\% \\
4 &
  APP China &
  20.63 &
  20090.9 &
  1328.0 &
  \multicolumn{1}{c|}{1263.6} &
  355.8 &
  \multicolumn{1}{c|}{380.5} &
  387.4 &
  \multicolumn{1}{c|}{271.7} &
  10.31\% &
  9.54\% \\
5 &
  Sougou &
  30.52 &
  13854.8 &
  990.1 &
  \multicolumn{1}{c|}{-} &
  444.9 &
  \multicolumn{1}{c|}{-} &
  393.2 &
  \multicolumn{1}{c|}{-} &
  13.20\% &
  - \\
6 &
  Meizu Market &
  25.29 &
  21875.8 &
  1870.7 &
  \multicolumn{1}{c|}{1406.2} &
  369.8 &
  \multicolumn{1}{c|}{398.1} &
  396.1 &
  \multicolumn{1}{c|}{330.1} &
  12.05\% &
  9.76\% \\
7 &
  AnZhi &
  23.86 &
  17345.3 &
  1013.0 &
  \multicolumn{1}{c|}{1596.1} &
  414.3 &
  \multicolumn{1}{c|}{398.5} &
  432.7 &
  \multicolumn{1}{c|}{352.4} &
  10.72\% &
  13.53\% \\ \hline
\end{tabular}
\begin{tablenotes}
    \item \small*: E and P represent the Ethereum Rinkeby environment and the layer-2 Polygon network, respectively.
    \item \small-: this market is no longer available in 2022.
\end{tablenotes}
\end{threeparttable}
}
\end{adjustbox}
\end{table*}

Table~\ref{tab:performance} lists the average results of each tested market for the Ethereum Rinkeby environment and the layer-2 Polygon network.
Note that the normal downloading time simply relies on the network quality between app markets and our AWS server instead of APK file sizes.
On top of the normal downloading time, \name introduces three steps of additional processing times, including
(i) about one second (or 1s) for extracting and validating checksums;
(ii) $\sim$0.4s for performing repackaging checks; 
and (iii) $\sim$0.4s for uploading apps to IPFS.
With these, the overall overhead introduced by \name is from 2.16\% to 20.56\%, with the median and average of 12.05\% and 11.39\%, respectively. 
We thus conclude that \name's performance overhead is around 12\%, a reasonable performance cost for a blockchain-based system.

\subsection{Gas Costs}
\label{sec:gas}

Throughout the 200 performance tests, we also collected the corresponding gas fees in Ether (for the Ethereum Rinkeby environment) and in Matic (for the layer-2 Polygon network). 
As of June 2023, the value of one Ether is approximately 1,860 USD, while one Matic is around 0.7 USD.

\begin{figure*}[t!]
	\centering
	\begin{subfigure}{0.48\textwidth}
		\begin{adjustbox}{center}
			\includegraphics[width=\linewidth]{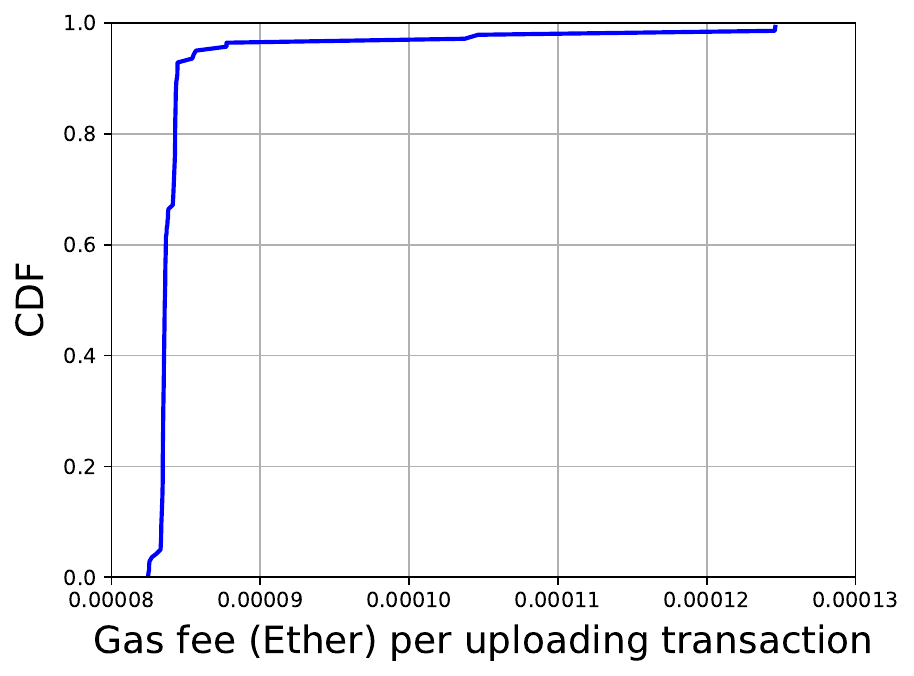}
		\end{adjustbox}
		\caption{In Ethereum (1 Ether$\approx$1.86K USD)}
		\label{fig:gasether}
	\end{subfigure}
	\hfill
	\begin{subfigure}{0.48\textwidth}
		\begin{adjustbox}{center}
			\includegraphics[width=\linewidth]{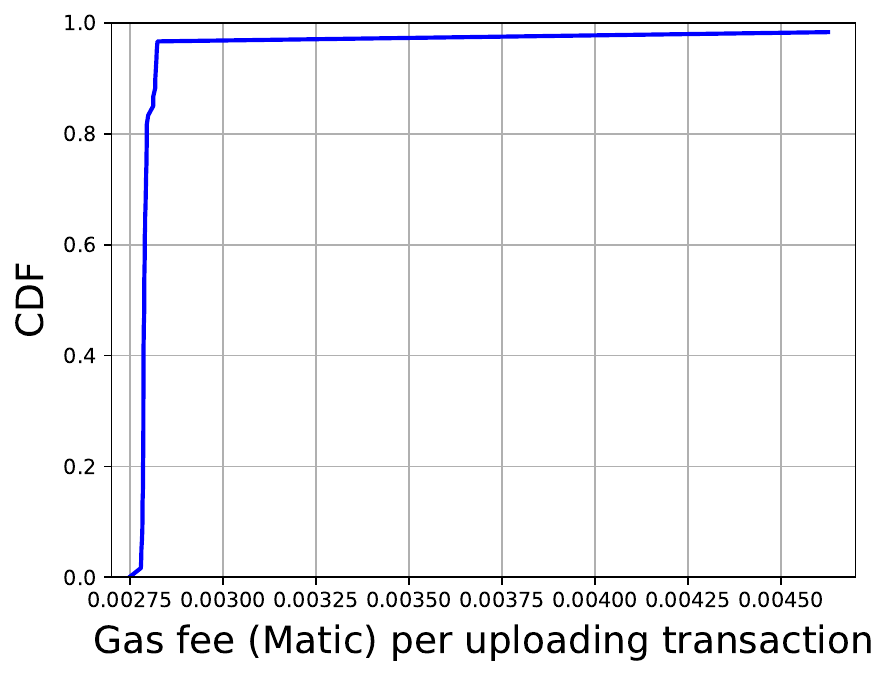}
		\end{adjustbox}
		\caption{In Polygon (1 Matic$\approx$0.7 USD)}
		\label{fig:gasmatic}
	\end{subfigure}
	\caption{CDF plots of gas fees per app uploading in \name.}
	\label{fig:GasFee}
\end{figure*}

\myfig\ref{fig:GasFee} shows the CDF (cumulative distribution function) plot per app uploading in \name, for both Ethereum and Polygon.
We can see that for the Ethereum Rinkeby environment, over 95\% gas fees are in the range of 0.00008245 Ether (0.1534 USD) and 0.00008773 Ether (0.1632 USD).
Only four tests consumed a gas fee over 0.0001 Ether, ranging from 0.00010374 and 0.00012461 Ether.
The average of all the gas fees of \name in the Ethereum Rinkeby environment is 0.00008466 Ether (0.1575 USD).
In contrast, the gas fees of \name in the Polygon network is even much smaller, with an average of only 0.002821 Matic (0.002 USD). 
Since only uploads in \name cost gas and one upload can serve all future downloads, we believe that such gas costs are quite acceptable for real-world deployment.

\subsection{Security}
\label{sec:security}

Since there are no real-world attacks against \name, we mimic a MITM (Man-In-The-Middle) attack and a repackaging attack to demonstrate \name's security effectiveness.

\begin{figure*}[t!]
\begin{adjustbox}{center}
\includegraphics[width=1.0\textwidth]{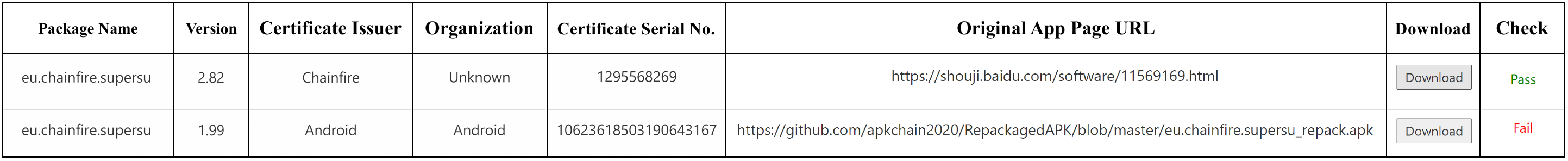}
\end{adjustbox}
\caption{A screenshot (with the surrounding table) to demonstrate a repackaged app successfully detected by \name.}
\label{fig:repackDemo}
\end{figure*}

\begin{figure}[t!]
\begin{adjustbox}{center}
\includegraphics[width=0.58\textwidth]{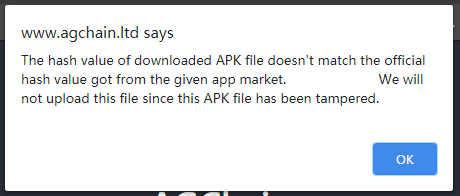}
\end{adjustbox}
\caption{A screenshot of \name defending against a MITM attack.}
\label{fig:MITMdemo}
\end{figure}

\textbf{Preventing a MITM attack.}
To mimic the MITM attack, we originally tried to control the network traffic of our AWS server using the mitmproxy tool~\cite{MITM} (since we cannot redirect app markets' traffic as real adversaries).
However, it turned out that AWS disallows this.
Hence, we have to redirect the target app page URL directly in our server code.
Specifically, when a user inputs this particular Tudou Video URL (\small\url{http://www.appchina.com/app/com.tudou.android}\normalsize) in \name, the actual app download URL will become the URL of a similar yet fake app (\small\url{https://github.com/apkchain2020/RepackagedAPK/blob/master/com.tudouship.android_4159.apk}\normalsize) we prepared.
During this process, \name finds that the MD5 retrieved from the original app page is f5580d6a58bb9d97c27929f1a9c585f1 while the MD5 calculated from the downloaded APK file is a05e5187-f4e9eb434bc3bbd792e35c54.  
Since these two checksums are different, \name shows an alert like \myfig~\ref{fig:MITMdemo} to the user, and does not allow this app to be uploaded.

\textbf{Detecting a repackaged app.}
To mimic the repackaging attack, we first select a target repackaged app that has ground truth as in \cite{Repack19} and also appears in Chinese app markets.
Our choice is the SuperSu rooting app on the Baidu market (\small\url{https://shouji.baidu.com/software/11569169.html}\normalsize).
We then upload this original app to \name and find that it ``passes'' the repackaging check, as shown in \myfig~\ref{fig:repackDemo}.
We further upload the repackaged app via the URL of \small\url{https://github.com/apkchain2020/RepackagedAPK/blob/master/eu.chainfire.supersu_repack.apk}\normalsize.
This time \name finds that the certificate serial number is different from that in our pre-generated certificate ID database using apps.
Indeed, \myfig~\ref{fig:repackDemo} also shows that the two serial numbers are different.
Hence, for the repackaged app, it ``fails'' the check.


\subsection{Decentralization}
\label{sec:decentralization}

During the 200 performance tests in different timings, we also identify IPFS gateways in over 20 different locations worldwide, which demonstrate the decentralization of \name.

Table~\ref{tab:IPFSgateways} provides a comprehensive list of the identified IPFS gateways.
We can see that these gateways are distributed across 21 different locations around the globe.
Around a half, ten, gateways are located in the United States.
This is probably because many IPFS nodes run in cloud servers, which are mainly provided by US companies.
Other than US, Europe holds seven gateways out of the remaining 11.
Compared with US and Europe, there are only three gateways in Asia (with the remaining gateway in Canada).
However, we anticipate that as FinTech gains popularity in Asia~\cite{AsiaFinTech20}, more IPFS nodes will be deployed locally, resulting in faster connections. 
Additionally, according to the RTT result in Table~\ref{tab:IPFSgateways}, nearby IPFS gateways usually have shorter RTTs, which suggests the value of identifying the fast IPFS gateways in \name for efficient app downloading (see \mysec\ref{sec:ipfsdesign}).

\begin{table}[t!]
\caption{The IPFS gateways identified in 21 different locations.}
\label{tab:IPFSgateways}
\begin{adjustbox}{center} 
\scalebox{0.8}{
\begin{tabular}{cccc}

\toprule
Gateway Domain          & IP Address        & Location                 & RTT (s)      \\ 

\midrule
ipfs.jbb.one            & 47.52.139.252     & Hong Kong SAR            & 0.04     \\
ipfs.smartsignature.io  & 13.231.230.12     & Tokyo, Japan             & 0.07     \\

10.via0.com             & 104.27.129.45     & San Francisco, U.S.A     & 0.13     \\
ipfs.kavin.rocks        & 104.28.5.229      & Dallas, U.S.A            & 0.14     \\

ipfs.runfission.com     & 34.233.130.24     & Ashburn, U.S.A           & 0.25     \\
ipfs.k1ic.com           & 39.101.143.85     & Beijing, China           & 0.51     \\

ipfs.2read.net          & 195.201.149.81    & Gunzenhausen, DE    & 0.55     \\
ipfs.drink.cafe         &  98.126.159.6     & Orange, U.S.A            & 0.56     \\
gateway.pinata.cloud    & 165.227.144.202   & Frankfurt, Germany  & 0.56  \\
ipfs.telos.miami        & 138.68.29.104     & Santa Clara, U.S.A       & 0.57     \\

hardbin.com             & 174.138.8.194     & Amsterdam, NL   & 0.57     \\
ipfs.fleek.co           & 44.240.5.243      & Portland, U.S.A          & 0.57     \\
ipfs.greyh.at           & 35.208.63.54      & Council Bluffs, U.S.A    & 0.69     \\
gateway.temporal.cloud  & 207.6.222.55      & Surrey, Canada           & 0.70     \\

ipfs.azurewebsites.net  & 13.66.138.105     & Redmond, U.S.A           & 0.72     \\
ipfs.best-practice.se   & 193.11.118.5      & Eskilstuna, Sweden       & 0.73     \\
ipfs.overpi.com         & 66.228.43.184     & Cedar Knolls, U.S.A.     & 0.74     \\
jorropo.net             & 163.172.31.60     & Paris, France            & 0.76     \\
jorropo.ovh             & 51.75.127.200     & Roubaix, France          & 0.76     \\

ipfs.stibarc.com        & 74.140.55.163     & Delaware, U.S.A          & 0.81     \\
ipfs.sloppyta.co        & 51.68.154.205     & Warsaw, Poland           & 0.83     \\

\bottomrule

\end{tabular}
}
\end{adjustbox}
\end{table}

\section{Discussion}
\label{sec:discuss}

In this section, we discuss several potential improvements \name could integrate with in the future and a few other potential extensions.


\textbf{Inviting developers.}
Although \name currently provides real-time app delegation for users to download apps just in time, it would be even more beneficial if \name could store a large number of apps in advance. 
This would greatly enhance user-friendliness. 
In order to achieve this, we propose inviting developers on Google Play to submit their apps to \name, thereby allowing their apps to reach a wider audience of users seeking decentralized access.
To implement this strategy, we can utilize the developer's email information listed on each Google Play app page. 
By crawling this information, we can send out invitation emails to developers, encouraging them to join \name. 
Furthermore, to incentivize developers to participate, we can cover the uploading gas fees for the first 10,000 developers who join our program and assist them in automatically uploading their apps.

\textbf{More secure.}
Although \name has implemented repackaging checks to detect repackaged apps, it currently lacks the ability to detect general Android malware. 
To address this limitation, we have devised a plan to integrate the malware scan feature of VirusTotal~\cite{VirusTotal}. 
Our approach involves utilizing the VirusTotal APIs to perform scans on each uploaded app on the server side. 
The results of these scans will be saved as a part of the app metadata in our smart contract, along with the URL to access the scan reports.
By incorporating this solution, users will be able to review the scan reports for apps during the downloading process. 
Additionally, we will provide explicit warnings for suspicious apps to further enhance user security and trust.

\textbf{Other potential extensions.}
In this paper, we implemented \name on the current Ethereum ecosystem, but it could be further extended in terms of supporting legacy blockchains~\cite{Bitcontracts21} and concurrent smart contract execution~\cite{DBLP:conf/ccs/WustMEKC20}.
Moreover, our crowdsourced server nodes could be enhanced by integrating TEE SGX~\cite{PrivacyGuar20} and becoming more serverless~\cite{Themis22}.
Additionally, the security and upgradeability of \name's smart contract could be protected by some of the state-of-the-art Ethereum protection techniques~\cite{SMARTSHIELD20, EVMPatch21, DBLP:journals/corr/abs-2206-00716}.


\section{Related Work}
\label{sec:related}

In this section, we present some other works that are closely related to \name.

\name is mostly related to several recent works~\cite{IoTDataBlockchain17, PubChain19, DClaims19, BBox20, DBLP:journals/jpdc/HeFPCTCPNX22} that also leveraged the blockchain and IPFS technology to construct decentralized systems in various domains.
For example, PubChain~\cite{PubChain19} is a decentralized publication platform that stored paper metadata in the blockchain layer and raw paper files in IPFS.
It introduced an incentive mechanism called PubCoin, which rewarded participants through a process referred to as ``publishing or reviewing as mining''.
Another paper on arXiv, DClaims~\cite{DClaims19}, presented a censorship-resistant service that utilized decentralized web notations to disseminate information on the Internet.
Similar to \name, it used Ethereum as the blockchain platform and integrated IPFS as the backend data storage.
Considering the frequency of web notations among numerous users, DClaims established a small network of nodes to aggregate multiple blockchain transactions and broadcast them collectively, thereby reducing the average transaction cost.
Apart from these works, Shafagh et al.~\cite{IoTDataBlockchain17} published a pioneer study on integrating blockchain and IPFS, drawing inspiration from the four-layer design of the Blockstack~\cite{Blockstack16} system: blockchain, virtualchain, routing, and storage layers.
In comparison to these related studies, \name stands out due to its unique characteristics in the following four aspects:


Firstly, unlike other blockchain systems that require users to choose between existing IT infrastructure and their own, our goal with \name was not to replace existing app markets. 
Instead, we designed \name as a gateway that not only provides permanent app delegation to end users but also leverages the vast number of apps available in existing markets. 
We believe that this design approach offers a unique opportunity to combine the benefits of traditional IT infrastructure with decentralized blockchain technology.

Secondly, we dramatically reduced gas costs in \name by proposing a set of design-level mechanisms.
In contrast, no aforementioned related works~\cite{IoTDataBlockchain17, PubChain19, DClaims19, BBox20} tried to do that. 
While there have been other works focusing on gas optimization, they primarily focused on language-level optimizations.
Specifically, GASPER~\cite{GASPER17} identified gas-costly smart contract coding patterns and summarized them into two categories, loop-related and useless codes.
MadMax~\cite{MadMax18} leveraged control- and data-flow analysis of smart contracts' bytecode to detect the gas-related vulnerabilities, including unbounded mass operations, non-isolated external calls, and integer overflows.
Additionally, GASOL~\cite{GASOL20} introduced a gas optimization approach by replacing multiple accesses to the global storage data with several single accesses to the data in local memory.
Accessing local memory incurs significantly fewer gas costs (3 Gas per access) compared to accessing storage data (each write access costing 20 Gas in the worst case and 5 Gas in the best case). 
However, these optimizations are still limited to the code level and are specifically designed for particular gas-costly patterns.

Thirdly, to the best of our knowledge, none of the existing studies on IPFS~\cite{IoTDataBlockchain17, PubChain19, DClaims19, BBox20, DFS20} explicitly mentioned the undistributed problem of IPFS, where IPFS files are cached in other nodes only when the node requests that file.
In addition to our attempt to this problem in \mysec\ref{sec:ipfsdesign}, the IPFS designers themselves also tried to address this issue.
Recently, they have launched Filecoin~\cite{Filecoin}, which is an incentive token mechanism aimed at encouraging peers in the IPFS network to remain online and take responsibility for storing files.
However, Filecoin requires users to pay for file storage, and the process of storing a 1MB file currently takes five to ten minutes in the Filecoin network~\cite{Filecoin}.
Due to these limitations, we built our own IPFS consortium network that leveraged IPFS gateways and crowdsourced server nodes to periodically backup files.

Lastly, we encountered three context-specific challenges that are unique to \name, as explained in \mysec\ref{sec:sustain}.

\section{Conclusion}
\label{sec:conclude}


In this paper, we proposed \name, a blockchain-based gateway for trustworthy -- permanent, distributed, and secure -- app delegation from existing app markets.
We addressed challenges in significantly reducing smart contract gas costs and enabling fully distributed IPFS-based file storage.
Additionally, we resolved three specific system issues for security and sustainability.
We implemented an \name prototype on Ethereum and Polygon blockchains, evaluating its performance, gas costs, security, and decentralization.
The results showed a 12\% performance overhead and a cost of around 0.002 USD per app upload (no cost for app download).
Further improvements include making \name more developer-friendly and secure.

\balance
\bibliographystyle{elsarticle-num} 
\bibliography{main}






\end{document}